\documentclass[runningheads]{llncs}

\usepackage{eccv}
\usepackage{float}

\usepackage{eccvabbrv}

\usepackage{graphicx}
\usepackage{booktabs}
\usepackage{makecell}

\usepackage[accsupp]{axessibility}  

\usepackage{hyperref}

\makeatletter
\newcommand{\printfnsymbol}[1]{%
  \textsuperscript{\@fnsymbol{#1}}%
}
\makeatother

\usepackage{orcidlink}

\begin{document}

\title{AIM 2024 Challenge on Compressed Video Quality Assessment: Methods and Results} 

\titlerunning{AIM 2024 Challenge on Compressed VQA: Methods and Results}

\author{Maksim Smirnov\thanks{M.~Smirnov (maksim.smirnov@graphics.cs.msu.ru), A.~Gushchin (alexander.gushchin@graphics.cs.msu.ru), A.~Antsiferova(aantsiferova@graphics.cs.msu.ru), D.~Vatolin (dmitriy@graphics.cs.msu.ru), and R.~Timofte (radu.timofte@uni-wuerzburg.de) were the challenge organizers, while the other authors participated in the challenge. \cref{affilations} contains the authors’ teams and affiliations. AIM 2024 webpage: \url{https://www.cvlai.net/aim/2024/}} \and Aleksandr Gushchin\printfnsymbol{1} 
\and Anastasia Antsiferova\printfnsymbol{1} 
\and\\ Dmitry Vatolin\printfnsymbol{1}  
\and Radu Timofte\printfnsymbol{1} 
\and Ziheng Jia 
\and Zicheng Zhang 
\and Wei Sun 
\and Jiaying Qian 
\and Yuqin Cao 
\and Yinan Sun 
\and Yuxin Zhu 
\and Xiongkuo Min 
\and Guangtao Zhai 
\and Kanjar De 
\and Qing Luo 
\and Ao-Xiang Zhang 
\and Peng Zhang 
\and Haibo Lei 
\and Linyan Jiang 
\and Yaqing Li 
\and Wenhui Meng 
\and Zhenzhong Chen 
\and Zhengxue Cheng 
\and Jiahao Xiao 
\and Jun Xu 
\and Chenlong He 
\and Qi Zheng 
\and Ruoxi Zhu 
\and Min Li 
\and Yibo Fan 
\and Zhengzhong Tu}

\authorrunning{M.~Smirnov, A.~Gushchin, A.~Antsiferova, D.~Vatolin, R.~Timofte et al.}

\institute{}


\maketitle

\begin{abstract}

Video quality assessment (VQA) is a crucial task in the development of video compression standards, as it directly impacts the viewer experience. This paper presents the results of the Compressed Video Quality Assessment challenge, held in conjunction with the Advances in Image Manipulation (AIM) workshop at ECCV 2024. The challenge aimed to evaluate the performance of VQA methods on a diverse dataset of 459 videos, encoded with 14 codecs of various compression standards (AVC/H.264, HEVC/H.265, AV1, and VVC/H.266) and containing a comprehensive collection of compression artifacts. To measure the methods performance, we employed traditional correlation coefficients between their predictions and subjective scores, which were collected via large-scale crowdsourced pairwise human comparisons. For training purposes, participants were provided with the Compressed Video Quality Assessment Dataset (CVQAD), a previously developed dataset of 1022 videos. Up to 30 participating teams registered for the challenge, while we report the results of 6 teams, which submitted valid final solutions and code for reproducing the results. Moreover, we calculated and present the performance of state-of-the-art VQA methods on the developed dataset, providing a comprehensive benchmark for future research. The dataset, results, and online leaderboard are publicly available at \url{https://challenges.videoprocessing.ai/challenges/compressed-video-quality-assessment.html}.

    \keywords{Video Compression \and Quality Assessment \and Challenge}
\end{abstract}

\section{Introduction}

Video constitutes the most significant portion of global internet traffic, increasing the network load and highlighting the critical need for efficient video compression. Developing and comparing new video encoders heavily depend on quality measurement, while many contemporary compression standards utilize modern techniques, including machine learning and neural network-based approaches. However, traditional image and video quality assessment (I/V-QA) methods (or metrics), like PSNR and SSIM, do not consider complex artifacts from these advanced compression standards. Furthermore, more recent quality assessment methods often correlate poorly with subjective scores of real-world compressed videos, highlighting the need for more robust and accurate methods.

A major issue in evaluating quality metrics for video compression distortions is the widespread use of outdated datasets, as user-generated content (UGC) datasets \cite{ghadiyaram2017capture,sinno2018large,nuutinen2016cvd2014,hosu2017konstanz,wang2019youtube, ying2021patch} with authentic distortions have become increasingly popular, supplanting synthetic legacy datasets \cite{wang2017videoset,de2010h,wang2016mcl,bampis2017study,min2020study,lin2015mcl,paudyal2014study,keimel2012tum} . Furthermore, most increasingly datasets containing compressed videos and subjective scores employ only H.264/AVC compression. The subjective scores for these videos often originate from in-lab tests, which involved a limited number of viewers and yielded only a few scores per video due to the complexity and high cost of subjective comparisons.

This AIM 2024 competition promotes innovations in compressed video quality assessment. We employed the methodology for this challenge from the MSU Video Quality Metrics Benchmark \cite{antsiferova2022video}. To create a diverse and comprehensive dataset for evaluation, we expanded the CVQAD dataset\cite{cvqad} with new videos, resulting in 459 videos for validation and testing (or public and private testing). We included videos with various compression artifacts to emulate real-life scenarios and evaluate submitted models fairly. The challenge consisted of two phases: development and testing. Twelve teams submitted their solutions, and six teams submitted their code for final validation.

This challenge is one of the AIM 2024 Workshop~\footnote{\url{https://www.cvlai.net/aim/2024/}} associated challenges on: sparse neural rendering~\cite{aim2024snr, aim2024snr_dataset}, UHD blind photo quality assessment~\cite{aim2024uhdbpqa}, compressed depth map super-resolution and restoration~\cite{aim2024cdmsrr}, raw burst alignment~\cite{aim2024rawburst}, efficient video super-resolution for AV1 compressed content~\cite{aim2024evsr}, video super-resolution quality assessment~\cite{aim2024vsrqa}, and video saliency prediction~\cite{aim2024vsp}.

\section{Challenge}

\subsection{Challenge datasets}

The training set employed in this study was sourced from the CVQAD dataset \cite{cvqad}, which comprises videos exhibiting a diverse range of compression artifacts. A new set of videos was collected using a similar methodology for validation and testing.

To minimize artifacts, 1080p reference videos were selected from a set of high-bitrate, open-source videos available on \url{www.vimeo.com}. A comprehensive search strategy was employed, incorporating a range of minor keywords to ensure maximum coverage of potential results. Only videos licensed under CC BY and CC0, with a minimum bitrate of 20 Mbps, were considered. The selected videos were then converted to YUV 4:2:0 chroma subsampling. A space-time-complexity clustering approach was employed to obtain a representative distribution of video complexity. Spatial complexity was calculated as the average size of x264-encoded I-frames normalized to the uncompressed frame size. In contrast, temporal complexity was calculated as the average P-frame size divided by the average I-frame size. The K-means algorithm \cite{lloyd1982least} divided the video collection into 36 clusters for CVQAD and 11 clusters for the new sets. Up to 10 candidate videos were randomly selected from each cluster, and a single video was manually chosen for inclusion in the final dataset. This process aimed to ensure a variety of genres, including sports, gaming, nature, interviews, and user-generated content. The resulting dataset consisted of 36 FullHD videos from CVQAD and 11 new FullHD videos for validation and test sets.

The videos were compressed using multiple encoders, including AVC/H.264, HEVC/H.265, AV1, and VVC/H.266 standards, to generate a diverse range of coding artifacts. Three presets were used for many encoders to increase the diversity of artifacts: the fastest provided a 30 FPS encoding speed, the medium provided 5 FPS, and the slowest provided a 1 FPS speed and higher quality. Not all videos underwent compression using all encoders. We compressed each video with three target bitrates — 1,000 kbps, 2,000 kbps, and 4,000 kbps — using a VBR mode (for encoders that support it) or with corresponding QP/CRF values that produce these bitrates. Primary streaming-video services recommend at most 4,500–8,000 kbps for FullHD encoding \cite{IBM_recommended_bitrates, Twitch_recommended_bitrates, YouTube_recommended_bitrates}. We avoided higher target bitrates because visible compression artifacts become almost unnoticeable, hindering subjective comparisons.

Ground-truth subjective scores were collected for the video dataset using the \url{www.subjectify.us} crowdsourcing platform. This platform employs a Bradley-Terry model to transform pairwise voting results into a score for each video.
The dataset was partitioned into subsets based on reference videos to mitigate the exponential growth of pairwise comparisons with the number of reference videos. Each subset consisted of one reference video and its corresponding compressed versions. Pairs were formed only for videos within the same subset, ensuring that only videos compressed from the same source were compared simultaneously. The comparison set also included the source videos themselves. Participants were presented with pairs of videos in full-screen mode, which they viewed sequentially. They were asked to select the video with the best visual quality or indicate that the two videos were of equivalent quality. An option to replay the videos was also available. Each participant was required to complete 12 pairwise comparisons, including two verification questions with an obvious higher-quality option. Responses from participants who failed to answer the verification questions correctly were excluded from the analysis. To ensure the reliability of the results, a minimum of 10 responses was collected for each pair. 

In total, around 500K valid answers were gathered from 10,800 individuals. The Bradley-Terry model was applied to the table of pairwise ranks, yielding subjective scores consistent within each group of videos compressed from a single reference video. 

This methodology was employed two times: for CVQAD (which represented a training set) and to label the validation and testing sets. Thus, we collected 1,022 videos from 36 reference videos, compressed with 22 encoding settings (for the training set), and 459 videos from 11 reference videos, compressed with 14 encoding settings (for validation and testing). Reference videos did the partitioning between validation and testing sets: each reference video with its corresponding compressed videos was placed into one of these sets. 11 reference videos were partitioned between validation and testing sets using the spatial and temporal complexity described above. Thus, the validation set consists of 5 references and 210 distorted videos, while the testing set consists of 6 references and 249 distorted videos. 

\begin{figure*}[htbp]
    \centering
    \includegraphics[width=0.8\linewidth]{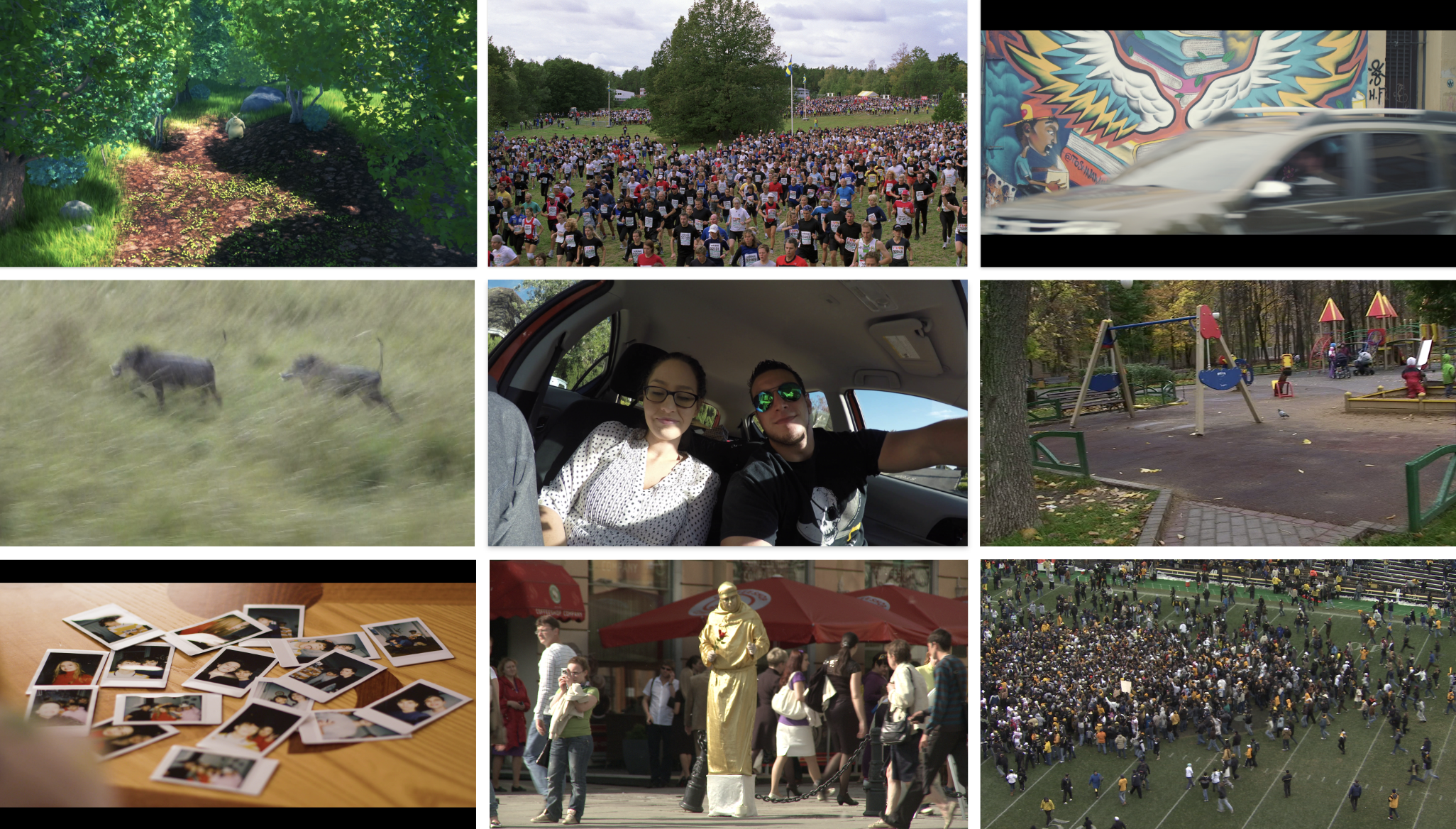}
    \caption{Samples from the videos in the CVQAC validation and test sets.
}
    \label{fig:my_diagram}
\end{figure*}

\subsection{Evaluation protocol}

This challenge employed a common evaluation approach to assess the performance of each method from two distinct perspectives: prediction monotonicity and prediction accuracy. We utilized the Spearman rank-order correlation coefficient (SROCC) and the Kendall rank-order correlation coefficient (KROCC) to enable the assessment of the monotonic relationship between the predicted and ground-truth subjective scores. Also, we employed the Pearson Linear Correlation coefficient (PLCC) to evaluate prediction accuracy. This metric measures the linear relationship between the predicted and ground-truth subjective scores.

The final score used for ranking was computed by averaging the SROCC, KROCC, and PLCC metrics. During the competition, scaled values of the results were presented to participants to encourage the improvement of their method's correlation values, which were knowingly overstated due to the chosen evaluation methodology.

The subjective scores were derived from pairwise comparisons of videos generated from the same original sequence and encoding preset. Consequently, these scores are only comparable within their respective groups, each consisting of three encoding bitrates times the number of codecs applied to the reference video. For each reference-video/preset pair, we calculated SROCC, KROCC, and PLCC between the quality assessment methods and subjective scores. To ensure statistically reliable results, we only considered groups with a sample size exceeding a predetermined threshold (15 for SROCC and 6 for KROCC and PLCC). We then applied the Fisher Z-transform \cite{fisher_transormation} (inverse hyperbolic tangent) to the results, weighted proportionally to group size, and averaged the transformed correlation coefficients. The inverse Fisher Z-transform yielded a single correlation coefficient for the entire dataset. The code example used for evaluation is available on GitHub: \url{https://github.com/msu-video-group/MSU_VQM_Compression_Benchmark}.

\subsection{Challenge phases}

The challenge dataset was partitioned into training, validation, and testing. The training set, comprising 1,022 videos with shared subjective scores, was made available to participants for method development. In contrast, the validation and testing sets, consisting of 210 and 249 videos, respectively, were used for evaluation purposes, and participants could not access their subjective scores. We also performed x265-lossless encoding of all compressed streams to simplify further evaluations.

During the development phase, participants could utilize any pre-trained or existing solutions without restrictions on model processing time or size. A testing system was established to facilitate method evaluation, allowing participants to submit their scores on the validation set and receive feedback on their performance, including their rank relative to other participants.

Two baseline models, MS-SSIM\cite{msssim} (Full-Reference, FR) and VSFA\cite{vsfa} (No-Reference, NR), were included to provide a reference point for participants. Both were implemented without ensemble methods or additional data. MS-SSIM is an advanced version of the traditional IQA method SSIM (Structural Similarity Index Measure) calculated over multiple scales using subsampling. VSFA integrates content-dependency and temporal-memory effects into a DNN, leveraging a pre-trained image content classification network and a GRU with a temporal pooling layer.

Throughout the development phase, 12 teams submitted their objective quality assessment methods, including one Full-Reference method. Upon the release of the testing set, participants were required to select one of their previous submissions and provide its validation and testing set scores, along with the corresponding code, checkpoints, and fact sheet. 

The final ranking was determined exclusively by performance on the test set and was announced after all participants had submitted their solutions and verified the reproducibility of their methods' predictions. Six teams provided fact sheets and source codes, which were used to establish the final ranking.

\section{Challenge results}

The challenge results are shown in \cref{tab:combined-results}. We only report the performances of the solutions, whose teams submitted their code and fact sheets. For IQA methods, we calculated frame-by-frame scores and averaged them to give the score for the whole video. The top three teams on the test set were TVQA-C, SJTU-MultimediaLab, and FudanVIP, each of which outperformed the VSFA baseline. The solution proposed by TVQA-C took the lead with a notable margin, although it was slightly outperformed in terms of SROCC by the SJTU-MultimediaLab solution, which contained nearly four times fewer parameters. Notably, all top-performing solutions leveraged features extracted from Visual-Language Models (VLMs). Furthermore, these solutions employed a multi-aspect approach to quality assessment, dividing methods into aesthetics (semantics) and technical quality (spatial information and artifacts), as well as local and global temporal information. The SJTU-MultimediaLab method also incorporated a module predicting the degree of video compression, which is crucial for the compressed video quality assessment. In contrast, the approach taken by the Test IQA team differed significantly from the others, utilizing interpretable classical analytic features and implementing a Full-Reference method. To facilitate a comprehensive comparison, we also report the results of other traditional and state-of-the-art methods on the collected test set, which are included in \cref{tab:sotas-results} for reference.

\begin{table}[tb]
  \caption{Results of AIM 2024 Compressed Video Quality Assessment Challenge. The ranking is based on the average values across 3 correlation coefficients. The number of model parameters is also included. Top-2 scores are shown in \textbf{bold} and \underline{underlined}.}
  \label{tab:combined-results}
  \tabcolsep=3.5pt
  \centering
  \scalebox{0.85}{%
\begin{tabular}{@{} l l c c c c c c }
  \toprule
  
  Rank & Team & Type & SROCC $^\uparrow$ & PLCC$^\uparrow$ & KROCC$^\uparrow$ & Result$^\uparrow$ & \#Params.(M) \\
  \midrule
  1 & TVQA-C                                       & NR Video &            \underline{0.9376}  &    \textbf{0.9772} &    \textbf{0.8505}  & \textbf{0.9218} & 388.34 \\
  
  2 & SJTU-MultimediaLab                                 & NR Video &    \textbf{0.9378} & \underline{0.9680} & \underline{0.8442} & \underline{0.9167} & 91.50 \\
  
 3 & FudanVIP                                  & NR Video & 0.9113 &           0.9568	  &            0.8009  & 0.8896 & 288 \\
 
  4 & Test IQA                                         & FR Image &           0.8873  &           0.9497	  &            0.7605  & 0.8658 & — \\
  5 & Fredlovematt                                  & NR Image &           0.8688  &            0.9411	 &            0.7617  & 0.8572 & 7 \\
6 & VPT                                                & NR Image &            0.8160  &           0.7741  &            0.5542	& 0.7148  & 317.3 \\ \hline
  & MS-SSIM\cite{msssim} (baseline) & FR Image  & 0.9149   &      0.9531  &            0.8062          & 0.8914 & \\
 & VSFA\cite{vsfa} (baseline)       & NR Video &          0.8844  &           0.9403  &         0.7913 & 0.8720 & \\ 
 
  \bottomrule
\end{tabular}}
\end{table}

\begin{table}[tb]
  \caption{Comparison of traditional and state-of-the-art methods performance on the collected test set. The ranking is based on the average values across 3 correlation coefficients. Top-2 scores are shown in \textbf{bold} and \underline{underlined}.}
  \label{tab:sotas-results}
  \centering
  \tabcolsep=5.5pt
  \scalebox{0.9}{%
\begin{tabular}{@{} l c  c  c  c  c }
  \toprule
  
  Method & Type & SROCC $^\uparrow$ & PLCC$^\uparrow$ & KROCC$^\uparrow$ & Result$^\uparrow$\\
  \midrule

BRISQUE\cite{brisque} & NR Image & 0.6303  & 0.6857 &  0.4923 & 0.6028 \\
WaDIQaM\cite{wadiqam} & NR Image & 0.6866 & 0.7054 & 0.5769 & 0.6563 \\
RankIQA\cite{rank} & NR Image & 0.7195 & 0.8365 & 0.5510 & 0.7023 \\
LPIPS\cite{lpips} & FR Image & 0.8370 & 0.9588 & 0.7022 & 0.8327 \\
Linearity\cite{linearity} & NR Image & 0.8756 & 0.9462 & 0.7592 & 0.8603\\
KonCept512\cite{koncept} & NR Image & 0.8703 & 0.9435 & 0.7827 & 0.8655 \\
DISTS\cite{dists} & NR Image & 0.8811 & 0.9612 & 0.7541 & 0.8655\\
VSFA\cite{vsfa} & NR Video & 0.8844 & 0.9403  & 0.7913 & 0.8720 \\ 
HaarPSI\cite{haar} & FR Image & 0.8897 & \underline{0.9770} & 0.7727 & 0.8798\\
PaQ-2-PiQ\cite{paq2piq} & NR Image & 0.9076 & 0.9529 & 0.8124 & 0.8910 \\
MS-SSIM\cite{msssim} & FR Image & 0.9149 &  0.9531  & 0.8062 & 0.8914 \\
CLIP-IQA+\cite{clip} & NR Image &  0.9023 &  0.9547 &  0.8184 & 0.8918 \\
DB-CNN\cite{dbcnn} & NR Image & 0.8502 & 0.9056 & \underline{0.9425} & 0.8994\\
MetaIQA\cite{meta} & NR Image & \textbf{0.9380} & 0.9640 & 0.8415 & 0.9145 \\
AHIQ\cite{ahiq} & FR Image &  \underline{0.9349} &  \textbf{0.9818} &  0.8385 & \underline{0.9184} \\
TOPIQ\cite{topiq} & NR Image & 0.9077 & 0.9446 & \textbf{0.9592} & \textbf{0.9372} \\

  \bottomrule
\end{tabular}}
\end{table}

\section{Challenge Teams and Solutions}
\subsection{TVQA-C}
\subsubsection{TVQA}
\begin{figure}
    \centering
    \includegraphics[width=1.0\textwidth]{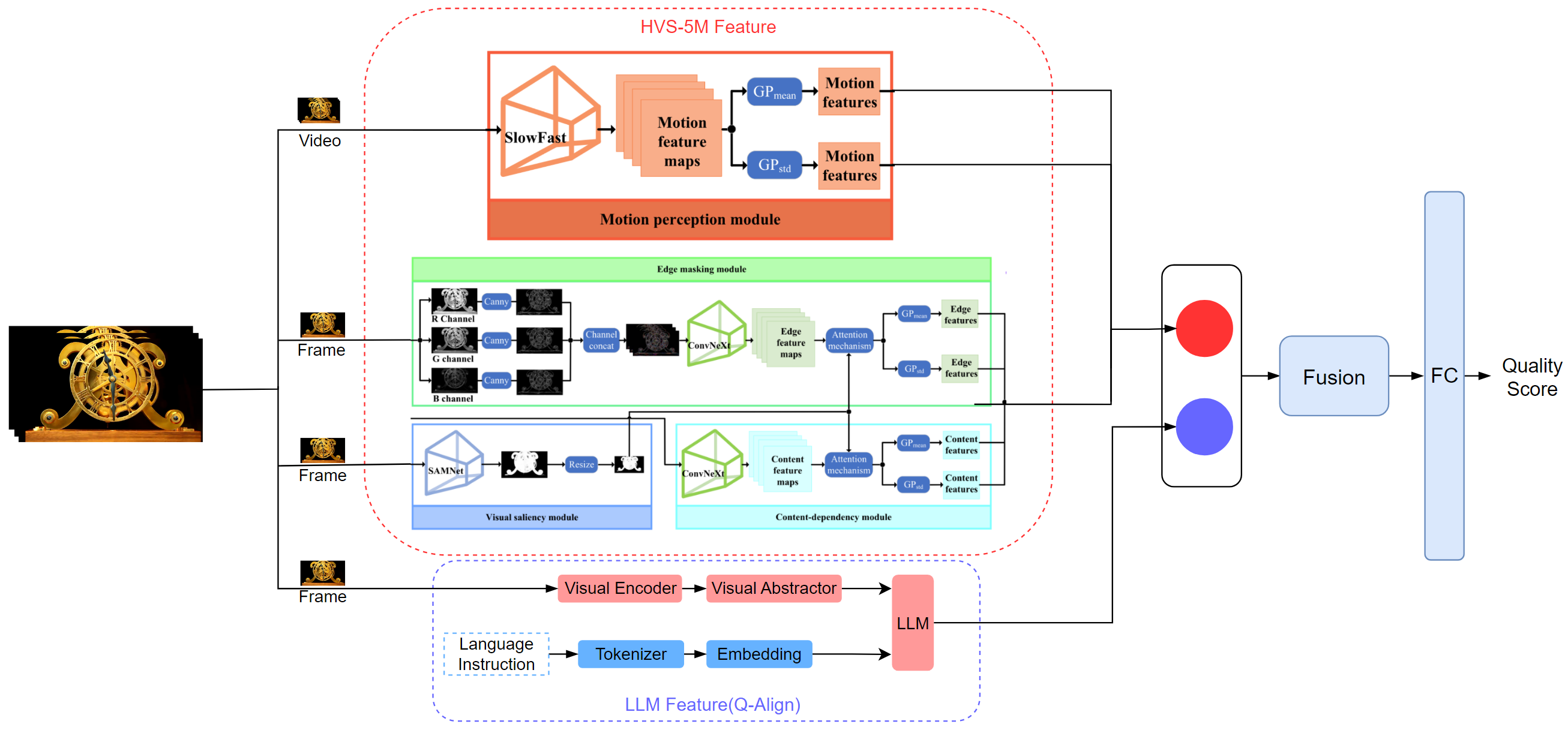}
    \caption{Architecture of TVQA, proposed by TVQA-C team.}
    \label{fig:team_name}
\end{figure}

We use HVS-5M~\cite{HVS_5M} to extract CNN-based video spatial features and motion features. Subsequently, Q-Align~\cite{wu2023qalign} is utilized to extract features from the video frame to enhance the semantic expression ability of the features. Then a feature fusion module is adopted to fuse the extracted features mentioned above. And finally, they are passed through a FC layer to obtain the quality score. The model architecture is shown in \ref{fig:team_name}.

During the training phase, PLCC loss and SROCC loss are used, and a KROCC loss is proposed to improve the performance of the model on KROCC metrics. In addition, since the same video has different scores in different groups, we introduce a group training strategy in TVQA-C.

\subsection{SJTU-MultimediaLab}
\subsubsection{Compression-RQ-VQA}
\begin{figure*}[h]
\centering
{\includegraphics[width=1\textwidth,height=0.7\textwidth]{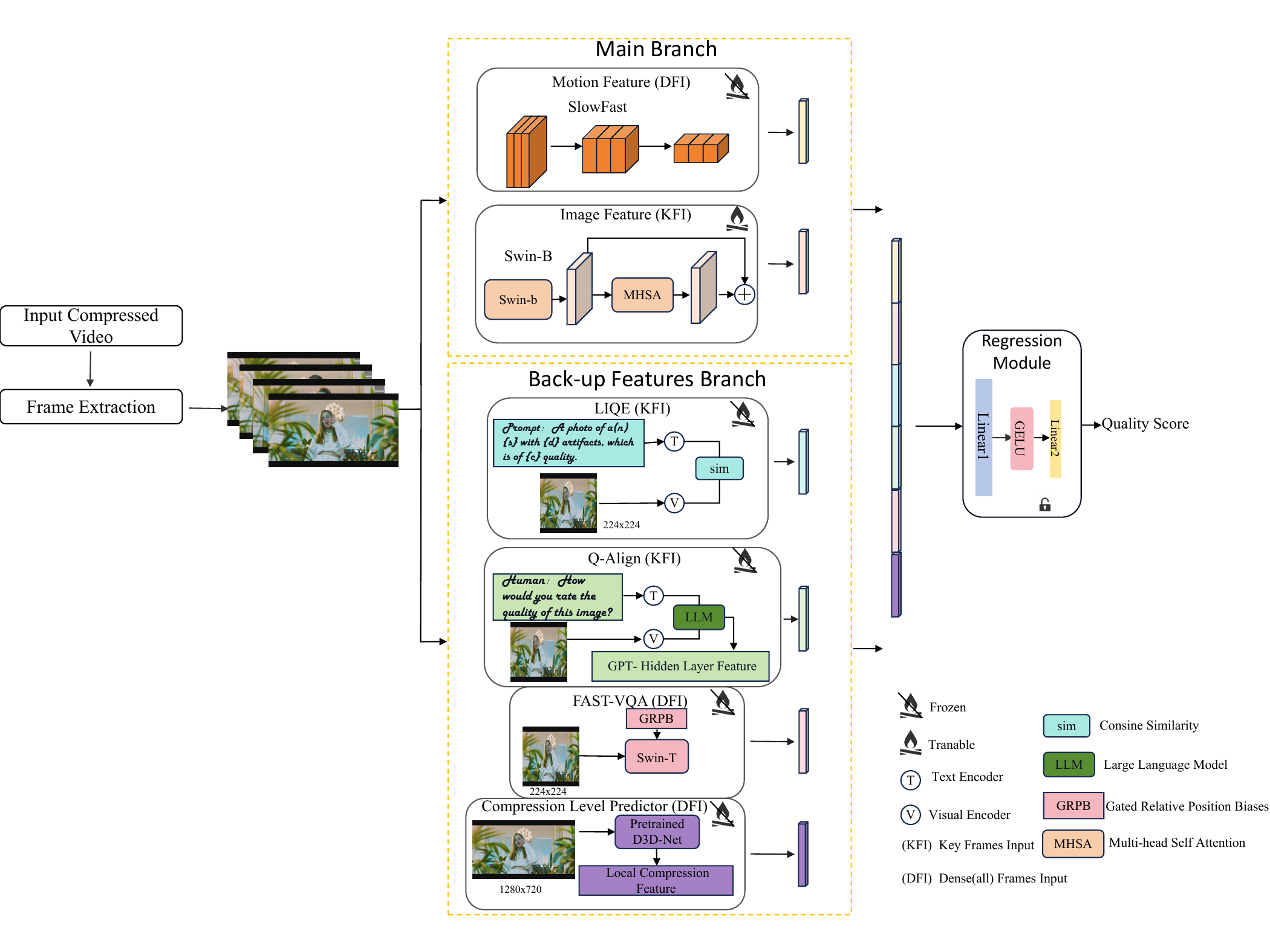}}
\caption{The overall structure of the Compression-RQ-VQA, proposed by SJTU-MultimediaLab, where KFI means key frames input (only extract one frame each second for input) and DFI means dense frames input (all the frames in the video are input).}
\label{fig-full}
\end{figure*}

This team proposes a blind (No-reference) compressed video quality assessment model based on the fusion of multiple perceptual quality features. The model architecture is primarily divided into two branches for extracting different perceptual quality features. The architecture of the proposed model is shown in Fig.\ref{fig-full}

In the main branch, motion features are extracted using the SlowFast\cite{feichtenhofer2019slowfast} model. The input consists of sequences of video frames divided into chunks per second. The output is the temporal average of the feature vectors sequence from the fast path of SlowFast model, resulting in a 256-dimensional feature vector. 

For the image perceptual features extraction, Swin-Transformer-b\cite{liu2021swin}  backbone is utilized. The input is a sequence of keyframes extracted uniformly from the video (one frame per second). After passing through the backbone, each frame's spatial token features are spatially averaged, resulting in a temporal feature vector sequence. The features of each keyframe are then temporally averaged to produce a 1024-dimensional feature vector of the whole video.

For the backup features branch, the LIQE feature extraction utilizes a pre-trained LIQE model \cite{zhang2023blind} to capture comprehensive perceptual features related to semantic information, distortion types, and overall quality perception of keyframes. The output keyframes LIQE features undergo temporal pooling to yield a 495-dimensional feature vector. Q-Align \cite{wu2023qalign} feature extraction employs the pretrained large multimodal language model (LMM), to extract generalized perceptual quality features of keyframes. While inputting keyframes, the output is obtained after removing the last projection layer of the LMM, followed by temporal pooling to produce a 4096-dimensional feature vector. Fast-VQA feature extraction aims to capture local quality perception features of the video. It utilizes the pre-trained Fast-VQA model \cite{wu2022fast} with the video as input, resulting in a 768-dimensional feature vector.

Compression feature extraction is a specialized module for evaluating the compressed videos quality. Following the work in \cite{wang2021rich}, it employs a D3D model \cite{stroud2020d3d} pre-trained on tasks related to predicting video compression levels. Inputting the entire video, it outputs a 1600-dimensional feature vector of compression information of the video local blocks. Finally, the extracted features from the model are concatenated and passed through a fully connected layer to obtain the predicted quality score of the video.

\bigbreak
\bigbreak

\subsection{FudanVIP}
\subsubsection{COVER}
\begin{figure*}[htbp]
    \centering
    \includegraphics[width=\textwidth]{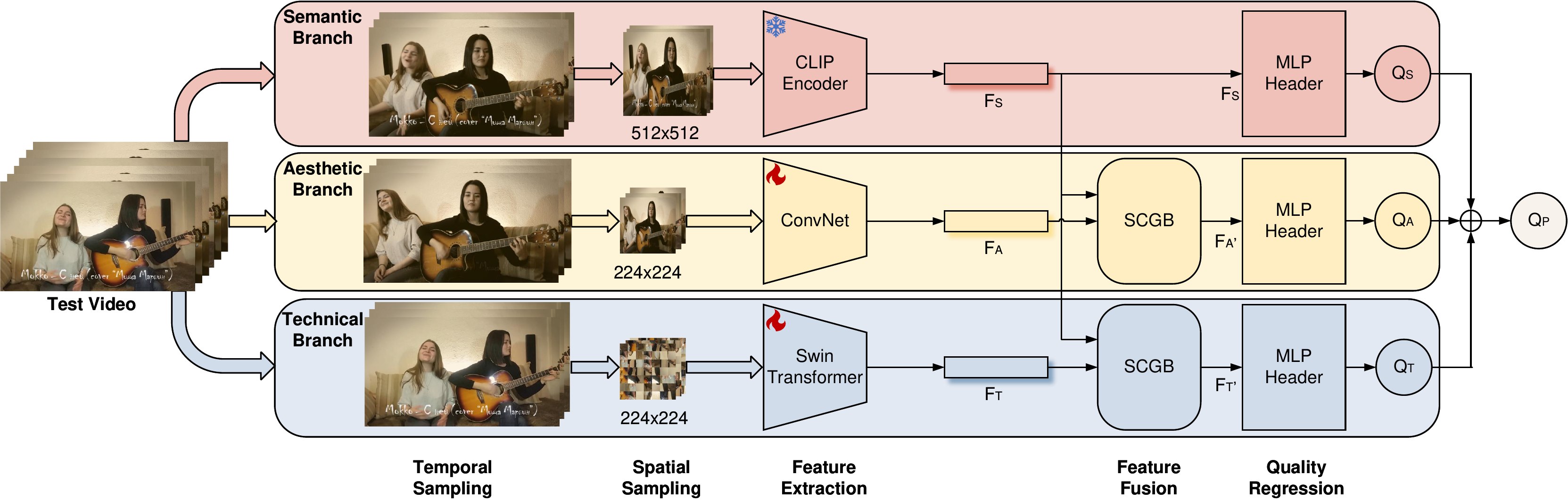}
    \caption{The architecture of  \textbf{\underline{C}}\textbf{\underline{O}}mprehensive \textbf{\underline{V}}ideo quality \textbf{\underline{E}}valuato\textbf{\underline{R}} (\textbf{COVER}), proposed by team FudanVIP.
COVER processes a video clip in three parallel branches: 1) a semantic branch that extracts high-level object-semantics-related information using a pre-trained CLIP image Encoder; 2) an aesthetic branch that leverages a ConvNet run on subsampled image thumbnails to analyze their looking; 3) a technical branch utilizing Swin Transformer to execute on fragments. The simplified cross-gating block (SCGB) is designed to fuse multi-branch features together, yielding the final quality score.
}
    \label{fig:my_diagram}
\end{figure*}

The network architecture of our proposed \textbf{\underline{C}}\textbf{\underline{O}}mprehensive \textbf{\underline{V}}ideo quality \textbf{\underline{E}}valuato\textbf{\underline{R}} (\textbf{COVER})~\cite{he2024cover} is illustrated in Fig.\ref{fig:my_diagram}. This network accepts videos that have been subjected to temporal-spatial sampling as its input. Its architecture is divided into three branches: a CLIP-based semantic branch, an aesthetic branch, and a technical branch, each consisting of a feature extraction module and a quality regression module. Notably, aesthetic and technical branches additionally incorporate a feature fusion module to integrate features from the semantic branch. The input video is processed through these branches to generate three scores, reflecting the video’s quality across the respective dimensions. The final score is the summary of scores from three dimensions.

\subsection{Test IQA}
\subsubsection{MMF Multimethod Fusion}
\begin{figure}[h!]
    \centering
    \includegraphics[width=0.8\linewidth]{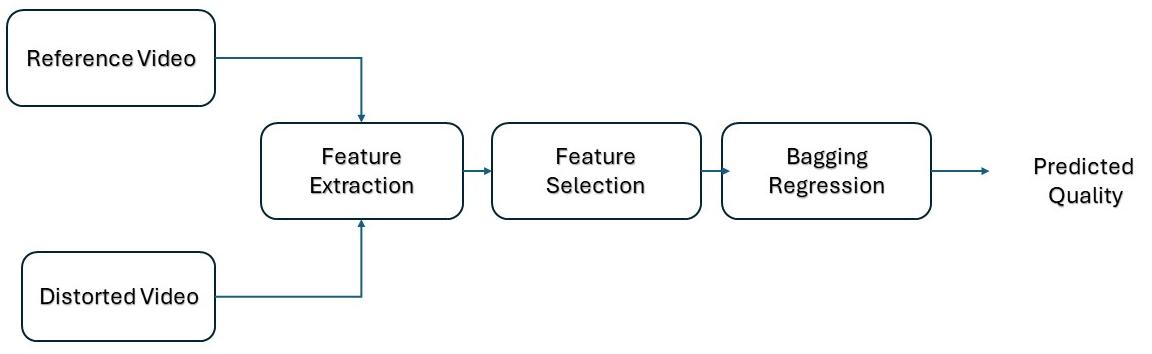}
    \caption{System Block Diagram for MMF Multimethod Fusion, proposed by Test IQA.}
    \label{fig:bd}
\end{figure}

The primary aim of this study is to investigate the concept of Multimethod Fusion within the Full Reference (FR) space. Our approach begins with non-learning-based features derived from transforms such as Discrete Cosine Transform (DCT) and Wavelet, on which we train a meta-model using the dataset. This approach is designed to create a tool that is highly customizable and offers some level of explainability. Given that different codecs operate based on distinct underlying principles, it is challenging for a single method to fit all scenarios. Therefore, our intention is to develop a flexible framework that initially employs several transform-based Image Quality Assessment (IQA) techniques and subsequently optimizes them across the dataset.
This research builds upon a previous paper of ours~\cite{de2018no}, with key modifications such as the transition from No-Reference (NR) to Full-Reference (FR) techniques, and the replacement of Random Forest with a Bagging Regression approach based on Support Vector Regression (SVR). The feature extraction stage is implemented in Python, utilizing the popular PyTorch Image Quality suite, PIQ~\cite{kastryulin2022piq,piq}. The list of features is presented in Tab.~\ref{tab:test_vqa}. To achieve optimal regression settings, we employed the Optuna hyperparameter optimization suite~\cite{optuna_2019}, aiming to minimize the Mean Squared Error (MSE) between predicted and ground-truth subjective scores. The overall system architecture is illustrated in Figure~\ref{fig:bd}

\begin{table}[htb]
\caption{Employed features in MMF Multimethod Fusion.}
\label{tab:test_vqa}
\begin{tabular}{lllll} \hline
IQA Features                                        & Abbreviation &  &  &  \\ \hline
Visual Saliency Induced Index                       & VSI          &  &  &  \\
Feature Similarity Index Measure                    & FSIM         &  &  &  \\
Haar Wavelet based Perceptual Similarity Index      & Haar         &  &  &  \\
Gradient Magnitude Similarity Deviation             & GMSD         &  &  &  \\
Multi Scale Gradient Magnitude Similarity Deviation & MS-GMSD      &  &  &  \\
Structural Similarity Index Measure                 & SSIM         &  &  &  \\
Multi Scale Structural Similarity Index Measure     & MS-SSIM      &  &  &  \\
DCT Subband Similarity Index                        & DSS          &  &  &  \\
Spectral Residual Based Similarity                  & SR-SIM       &  &  &  \\
Mean Deviation Similarity Index                     & MDSI         &  &  &  \\ \hline
\end{tabular}
\end{table}

\subsection{Fredlovematt}
\subsubsection{IQA-LLaVa}
Our solution leverages a pre-trained LLaVA-7b model \cite{liu2023llava,liu2024llavanext,liu2023improvedllava} fine-tuned using instruction tuning \cite{liu2023llava} on the q-instruct dataset \cite{wu2023qinstruct} and a preprocessed pseudo-labeled dataset provided by the competition organizers. The core steps of our image and video scoring tool are as follows:

\begin{itemize}
    \item \textbf{Data Preparation}: We fine-tuned the pre-trained LLaVA model using the q-instruct dataset and the preprocessed pseudo-labeled dataset.
    \item \textbf{Frame Extraction}: For video quality evaluation, we extract frames at regular intervals from the videos.
    \item \textbf{Image Quality Evaluation}: Each extracted frame is scored for quality using the fine-tuned LLaVA model.
    \item \textbf{Score Aggregation}: The scores for all frames are averaged to determine the overall video quality score.
\end{itemize}

The IQA-LLaVa pipeline diagram is shown in Fig.~\ref{fig:pipeline}.:

\begin{figure}[h]
    \centering
    \includegraphics[width=\textwidth]{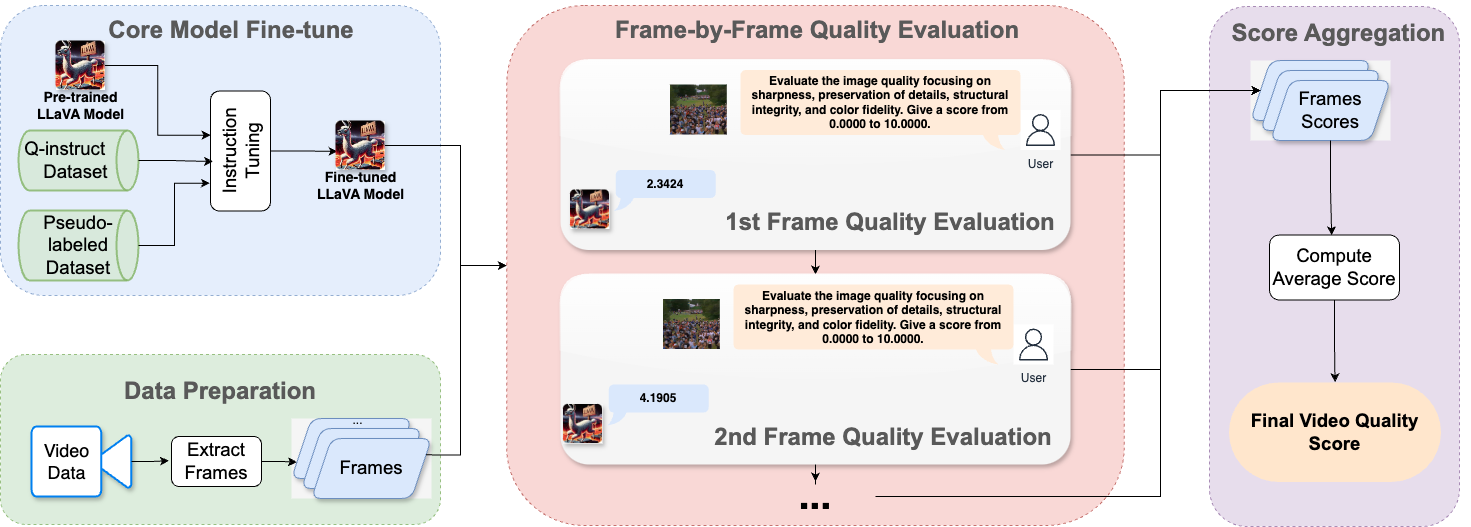}
    \caption{Diagram of IQA-LLaVa architecture, proposed by team Fredlovematt.}
    \label{fig:pipeline}
\end{figure}

\subsection{VPT}
\subsubsection{HIVE}
\begin{figure*}[htp]
    \centering
\includegraphics[width=0.98\textwidth]{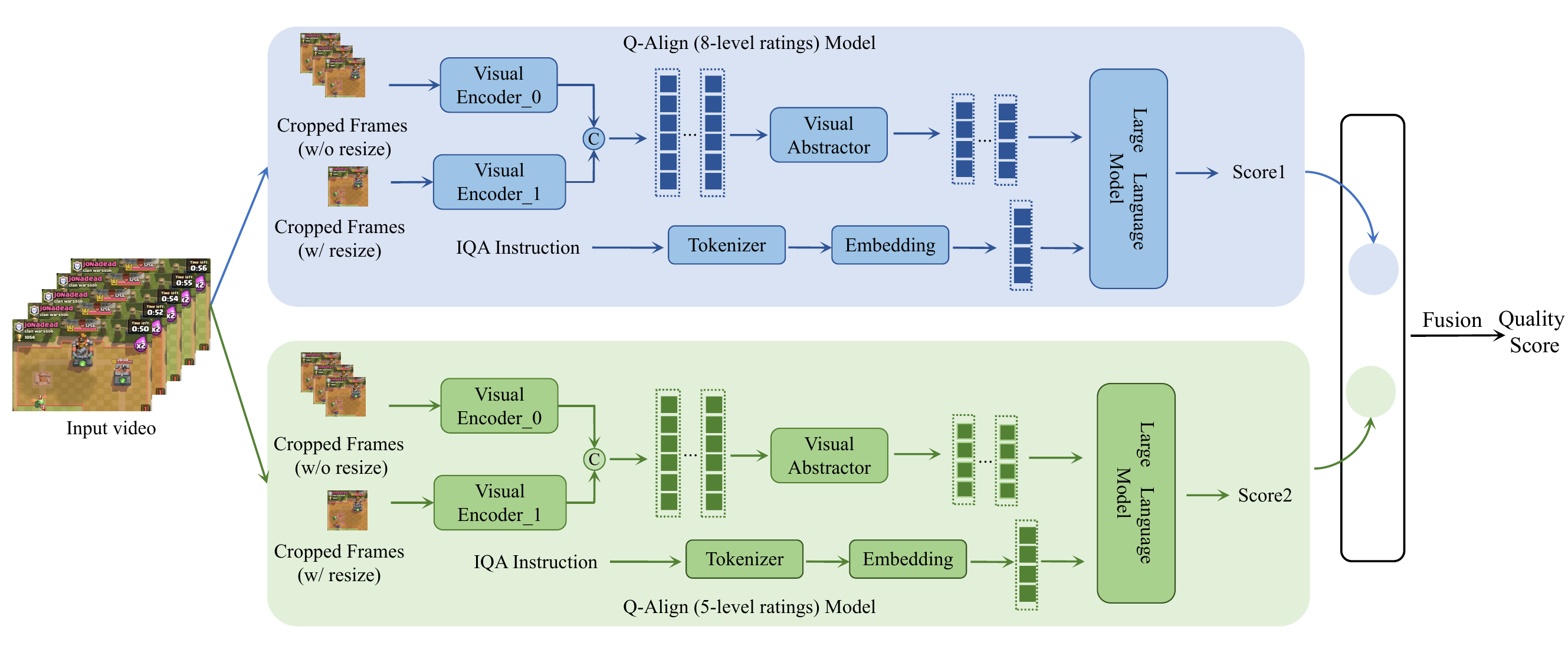}
    \caption{The network architecture of HIVE, proposed by team VPT.}
    \label{fig:framework}
\end{figure*}

\newcommand{\para}[1]{\vskip 2pt\noindent{\textit{\textbf{#1:\ }}}}
This team proposes the \underline{\textbf{Hi}}erarchical \underline{\textbf{V}}isual \underline{\textbf{E}}valuator (HIVE), a model based on large multi-modality models (LMMs) for VQA tasks. The overall framework is shown in Fig.~\ref{fig:framework}.
HIVE builds on Q-Align \cite{wu2023qalign} and introduces two key components to address the limitations of the original 5-level quality descriptors of Q-Align on VQA tasks. 
a) \textbf{Leveraging different levels of quality descriptions}: HIVE utilizes two LMMs based on Q-Align with different levels of quality descriptions, motivated by the noticeable quantization effect of the final predicted score when using a single Q-Align.
By adopting two Q-Align models with different quality descriptors to predict and fuse the scores, HIVE transforms the scores from 'coarse' to 'fine', resulting in more well-distributed scores.
b) \textbf{Developing dual-encoder for local-global fusion}: While the original Q-Align only processes the resized frames, HIVE develops a dual-encoder approach to handle both the original local patches and resized global patches separately. By fusing the features from these two perspectives, HIVE captures comprehensive information from local and global, leading to more accurate score predictions.

\para{Training Details}
The LMMs in HIVE are initialized with publicly pre-trained Q-Align and trained on NVIDIA Tesla A100 GPUs (40G) using the challenge dataset \cite{antsiferova2022video} with the initial learning rate of 2e-5 for several epochs independently.

\para{Testing Details}
HIVE processes a test video using the two LMMs and obtains the final predicted score by weighted averaging the two scores.

\section{Conclusion}

This paper has introduced the AIM 2024 challenge on compressed quality assessment, which aimed to evaluate the performance of quality assessment methods on compression artifacts. A comprehensive dataset of 459 videos for testing and 1024 videos for training was collected, featuring a range of compression standards and encoding settings. Six teams participated in the challenge, with team TVQA-C achieving the top ranking. This challenge has contributed to the advancement of compressed quality assessment research, and we anticipate further progress in this field as researchers continue to develop and refine their methods.

\section{Challenge Teams and Affiliations}\label{affilations}
\subsection{Organizers of AIM 2024 Compressed Video Quality Assessment Challenge}
\textit{\textbf{Members:}}\\
Maksim Smirnov$^{1,2,\dagger}$ (maksim.smirnov@graphics.cs.msu.ru), \\Aleksandr Gushchin$^{1,3,4,\dagger}$ (alexander.gushchin@graphics.cs.msu.ru), \\Anastasia Antsiferova$^{3,4,\dagger}$ (aantsiferova@graphics.cs.msu.ru),  \\Dmitry Vatolin$^{1,4}$ (dmitriy@graphics.cs.msu.ru), \\Radu Timofte$^{5}$ (radu.timofte@uni-wuerzburg.de)
\\
\textit{\textbf{Affiliations:}}\\
$^\dagger$ Equal contribution\\
$^1$ Lomonosov Moscow State University\\
$^2$ Yandex Research\\
$^3$ ISP RAS Research Center for Trusted Artificial Intelligence \\
$^4$ MSU Institute for Artificial Intelligence\\
$^5$ University of Würzburg, Germany

\subsection{TVQA-C}
\textit{\textbf{Members:}}\\
Qing Luo$^{*,1}$ (luoqing.94@qq.com), Ao-Xiang Zhang$^{*,1,2}$ (zax@e.gzhu.edu.cn), Peng Zhang$^1$, Haibo Lei$^1$, Linyan Jiang$^1$, Yaqing Li$^1$\\
$^*$: Equal contribution.
\\
\textit{\textbf{Affiliations:}}\\
$^1$: Tencent, China\\ 
$^2$: School of Computer Science
and Cyber Engineering, Guangzhou University, China\\ 

\subsection{SJTU-MultimediaLab}
\textit{\textbf{Members:}}\\
Ziheng Jia$^1$ (jzhws1@sjtu.edu.cn), Zicheng Zhang$^1$ (zzc1998@sjtu.edu.cn), Wei Sun$^{1}$ (sunguwei@sjtu.edu.cn), Jiaying Qian$^{1}$ (2022qjy@sjtu.edu.cn), Yuqin Cao$^{1}$ (caoyuqin@sjtu.edu.cn), Yinan Sun$^1$ (yinansun@sjtu.edu.cn), Yuxin Zhu$^1$(rye2000@sjtu.edu.cn), Xiongkuo Min$^{1}$ (minxiongkuo@sjtu.edu.cn), Guangtao Zhai$^{1}$ (zhaiguangtao@sjtu.edu.cn)\\
\textit{\textbf{Affiliations:}}\\
$^1$: Shanghai Jiao Tong University, China\\

\subsection{FudanVIP}
\textit{\textbf{Members:}}\\
Chenlong He$^1$, Qi Zheng$^1$, Ruoxi Zhu$^1$, Min Li$^1$, Yibo Fan$^1$, Zhengzhong Tu$^2$\\
\textit{\textbf{Affiliations:}}\\
$^1$: Fudan University, China\\
$^2$: University of Texas at Austin, America\\

\subsection{Test IQA}
\textit{\textbf{Members:}}\\
Kanjar De$^1$ (kanjar.de@gmail.com)\\
\textit{\textbf{Affiliations:}}\\
$^1$: ERCIM (Fraunhofer HHI), Germany\\

\subsection{Fredlovematt}
\textit{\textbf{Members:}}\\
Jiahao Xiao$^1$ (boss123456@sjtu.edu.cn), Jun Xu$^1$ (xujunzz@sjtu.edu.cn), Zhengxue Cheng$^1$ (zxcheng@sjtu.edu.cn)\\
\textit{\textbf{Affiliations:}}\\
$^1$: Shanghai Jiao Tong Univ. (China)\\

\subsection{VPT}
\textit{\textbf{Members:}}\\
Wenhui Meng$^1$ (whmeng@whu.edu.cn), Zhenzhong Chen$^1$ \\
\textit{\textbf{Affiliations:}}\\
$^1$: Wuhan University, China\\

\newpage

\section*{Acknowledgements}
This work was partially supported by the Humboldt Foundation. We thank the AIM 2024 sponsors: Meta Reality Labs, KuaiShou, Huawei, Sony Interactive Entertainment and University of W\"urzburg (Computer Vision Lab). 

The evaluations for this research were carried out using the MSU-270 supercomputer of Lomonosov Moscow State University.

%
%
\bibliographystyle{splncs04}
\bibliography{main}

\begin{thebibliography}{10}
\providecommand{\url}[1]{\texttt{#1}}
\providecommand{\urlprefix}{URL }
\providecommand{\doi}[1]{https://doi.org/#1}

\bibitem{IBM_recommended_bitrates}
Recommended encoding settings for ibm watson media. \url{https://support.video.ibm.com/hc/en-us/articles/207852117-Internet-connection-and-recommended-encoding-settings}, accessed: 2022-08-12

\bibitem{Twitch_recommended_bitrates}
Recommended encoding settings for twitch streaming. \url{https://stream.twitch.tv/encoding/}, accessed: 2022-08-12

\bibitem{YouTube_recommended_bitrates}
Recommended encoding settings for youtube. \url{https://support.google.com/youtube/answer/2853702}, accessed: 2022-08-12

\bibitem{optuna_2019}
Akiba, T., Sano, S., Yanase, T., Ohta, T., Koyama, M.: Optuna: A next-generation hyperparameter optimization framework. In: Proceedings of the 25th {ACM} {SIGKDD} International Conference on Knowledge Discovery and Data Mining (2019)

\bibitem{antsiferova2022video}
Antsiferova, A., Lavrushkin, S., Smirnov, M., Gushchin, A., Vatolin, D., Kulikov, D.: Video compression dataset and benchmark of learning-based video-quality metrics. Advances in Neural Information Processing Systems  \textbf{35},  13814--13825 (2022)

\bibitem{bampis2017study}
Bampis, C.G., Li, Z., Moorthy, A.K., Katsavounidis, I., Aaron, A., Bovik, A.C.: Study of temporal effects on subjective video quality of experience. IEEE Transactions on Image Processing  \textbf{26}(11),  5217--5231 (2017)

\bibitem{wadiqam}
Bosse, S., Maniry, D., Müller, K.R., Wiegand, T., Samek, W.: Deep neural networks for no-reference and full-reference image quality assessment. IEEE Transactions on Image Processing  \textbf{27}(1),  206--219 (2018). \doi{10.1109/TIP.2017.2760518}

\bibitem{topiq}
Chen, C., Mo, J., Hou, J., Wu, H., Liao, L., Sun, W., Yan, Q., Lin, W.: Topiq: A top-down approach from semantics to distortions for image quality assessment. arXiv preprint arXiv:2308.03060  (2023)

\bibitem{aim2024rawburst}
Conde, M.V., Bishop, T., Timote, R., Kolmet, M., MacEwan, D., Vinod, V., Tan, J., et~al.: {AIM} 2024 challenge on raw burst alignment via optical flow estimation. In: Proceedings of the European Conference on Computer Vision (ECCV) Workshops (2024)

\bibitem{aim2024evsr}
Conde, M.V., Lei, Z., Li, W., Katsavounidis, I., Timofte, R., et~al.: {AIM} 2024 challenge on efficient video super-resolution for av1 compressed content. In: Proceedings of the European Conference on Computer Vision (ECCV) Workshops (2024)

\bibitem{aim2024cdmsrr}
Conde, M.V., Vasluianu, F.A., Xiong, J., Ye, W., Ranjan, R., Timofte, R., et~al.: Compressed depth map super-resolution and restoration: {AIM} 2024 challenge results. In: Proceedings of the European Conference on Computer Vision (ECCV) Workshops (2024)

\bibitem{fisher_transormation}
Corey, D., Dunlap, W., Burke, M.: Averaging correlations: Expected values and bias in combined pearson rs and fisher's z transformations. Journal of General Psychology - J GEN PSYCHOL  \textbf{125},  245--261 (07 1998). \doi{10.1080/00221309809595548}

\bibitem{de2018no}
De, K., Masilamani, V.: No-reference image quality measure for images with multiple distortions using random forests for multi method fusion. Image Analysis and Stereology  \textbf{37}(2),  105--117 (2018)

\bibitem{de2010h}
De~Simone, F., Tagliasacchi, M., Naccari, M., Tubaro, S., Ebrahimi, T.: A h. 264/avc video database for the evaluation of quality metrics. In: 2010 IEEE International Conference on Acoustics, Speech and Signal Processing. pp. 2430--2433. IEEE (2010)

\bibitem{dists}
Ding, K., Ma, K., Wang, S., Simoncelli, E.: Image quality assessment: Unifying structure and texture similarity. IEEE Transactions on Pattern Analysis and Machine Intelligence  \textbf{PP}, ~1--1 (12 2020). \doi{10.1109/TPAMI.2020.3045810}

\bibitem{feichtenhofer2019slowfast}
Feichtenhofer, C., Fan, H., Malik, J., He, K.: Slowfast networks for video recognition. In: Proceedings of the IEEE/CVF international conference on computer vision. pp. 6202--6211 (2019)

\bibitem{ghadiyaram2017capture}
Ghadiyaram, D., Pan, J., Bovik, A.C., Moorthy, A.K., Panda, P., Yang, K.C.: In-capture mobile video distortions: A study of subjective behavior and objective algorithms. IEEE Transactions on Circuits and Systems for Video Technology  \textbf{28}(9),  2061--2077 (2017)

\bibitem{cvqad}
Gushchin, A., Smirnov, M., Antsiferova, A., Lyapustin, E., Vatolin, D.: Msu cvqad: Compressed video quality assessment dataset (2022), \url{https://videoprocessing.ai/datasets/cvqad.html}

\bibitem{he2024cover}
He, C., Zheng, Q., Zhu, R., Zeng, X., Fan, Y., Tu, Z.: Cover: A comprehensive video quality evaluator. In: Proceedings of the IEEE/CVF Conference on Computer Vision and Pattern Recognition. pp. 5799--5809 (2024)

\bibitem{aim2024uhdbpqa}
Hosu, V., Conde, M.V., Timofte, R., Agnolucci, L., Zadtootaghaj, S., Barman, N., et~al.: {AIM} 2024 challenge on uhd blind photo quality assessment. In: Proceedings of the European Conference on Computer Vision (ECCV) Workshops (2024)

\bibitem{hosu2017konstanz}
Hosu, V., Hahn, F., Jenadeleh, M., Lin, H., Men, H., Szir{\'a}nyi, T., Li, S., Saupe, D.: The konstanz natural video database (konvid-1k). In: 2017 Ninth international conference on quality of multimedia experience (QoMEX). pp.~1--6. IEEE (2017)

\bibitem{koncept}
Hosu, V., Lin, H., Sziranyi, T., Saupe, D.: Koniq-10k: An ecologically valid database for deep learning of blind image quality assessment. IEEE Transactions on Image Processing  \textbf{29},  4041--4056 (2020). \doi{10.1109/TIP.2020.2967829}

\bibitem{piq}
Kastryulin, S., Zakirov, D., Prokopenko, D.: {PyTorch Image Quality}: Metrics and measure for image quality assessment (2019), \url{https://github.com/photosynthesis-team/piq}, open-source software available at https://github.com/photosynthesis-team/piq

\bibitem{kastryulin2022piq}
Kastryulin, S., Zakirov, J., Prokopenko, D., Dylov, D.V.: Pytorch image quality: Metrics for image quality assessment (2022). \doi{10.48550/ARXIV.2208.14818}, \url{https://arxiv.org/abs/2208.14818}

\bibitem{keimel2012tum}
Keimel, C., Redl, A., Diepold, K.: The tum high definition video datasets. In: 2012 Fourth international workshop on quality of multimedia experience. pp. 97--102. IEEE (2012)

\bibitem{ahiq}
Lao, S., Gong, Y., Shi, S., Yang, S., Wu, T., Wang, J., Xia, W., Yang, Y.: Attentions help cnns see better: Attention-based hybrid image quality assessment network. arXiv preprint arXiv:2204.10485  (2022)

\bibitem{vsfa}
Li, D., Jiang, T., Jiang, M.: Quality assessment of in-the-wild videos. In: Proceedings of the 27th ACM International Conference on Multimedia. pp. 2351--2359 (2019)

\bibitem{linearity}
Li, D., Jiang, T., Jiang, M.: Norm-in-norm loss with faster convergence and better performance for image quality assessment. In: Proceedings of the 28th ACM International Conference on Multimedia. pp. 789--797 (2020)

\bibitem{lin2015mcl}
Lin, J.Y., Song, R., Wu, C.H., Liu, T., Wang, H., Kuo, C.C.J.: Mcl-v: A streaming video quality assessment database. Journal of Visual Communication and Image Representation  \textbf{30}, ~1--9 (2015)

\bibitem{liu2023improvedllava}
Liu, H., Li, C., Li, Y., Lee, Y.J.: Improved baselines with visual instruction tuning (2023)

\bibitem{liu2024llavanext}
Liu, H., Li, C., Li, Y., Li, B., Zhang, Y., Shen, S., Lee, Y.J.: Llava-next: Improved reasoning, ocr, and world knowledge (January 2024), \url{https://llava-vl.github.io/blog/2024-01-30-llava-next/}

\bibitem{liu2023llava}
Liu, H., Li, C., Wu, Q., Lee, Y.J.: Visual instruction tuning (2023)

\bibitem{rank}
Liu, X., Van De~Weijer, J., Bagdanov, A.D.: Rankiqa: Learning from rankings for no-reference image quality assessment. In: 2017 IEEE International Conference on Computer Vision (ICCV). pp. 1040--1049 (2017). \doi{10.1109/ICCV.2017.118}

\bibitem{liu2021swin}
Liu, Z., Lin, Y., Cao, Y., Hu, H., Wei, Y., Zhang, Z., Lin, S., Guo, B.: Swin transformer: Hierarchical vision transformer using shifted windows. In: Proceedings of the IEEE/CVF international conference on computer vision. pp. 10012--10022 (2021)

\bibitem{lloyd1982least}
Lloyd, S.: Least squares quantization in pcm. IEEE transactions on information theory  \textbf{28}(2),  129--137 (1982)

\bibitem{min2020study}
Min, X., Zhai, G., Zhou, J., Farias, M.C., Bovik, A.C.: Study of subjective and objective quality assessment of audio-visual signals. IEEE Transactions on Image Processing  \textbf{29},  6054--6068 (2020)

\bibitem{brisque}
Mittal, A., Moorthy, A.K., Bovik, A.C.: No-reference image quality assessment in the spatial domain. IEEE Transactions on Image Processing  \textbf{21}(12),  4695--4708 (2012). \doi{10.1109/TIP.2012.2214050}

\bibitem{aim2024vsrqa}
Molodetskikh, I., Borisov, A., Vatolin, D.S., Timofte, R., et~al.: {AIM} 2024 challenge on video super-resolution quality assessment: Methods and results. In: Proceedings of the European Conference on Computer Vision (ECCV) Workshops (2024)

\bibitem{aim2024vsp}
Moskalenko, A., Bryntsev, A., Vatolin, D.S., Timofte, R., et~al.: {AIM} 2024 challenge on video saliency prediction: Methods and results. In: Proceedings of the European Conference on Computer Vision (ECCV) Workshops (2024)

\bibitem{aim2024snr}
Nazarczuk, M., Catley-Chandar, S., Tanay, T., Shaw, R., Pérez-Pellitero, E., Timofte, R., et~al.: {AIM} 2024 sparse neural rendering challenge: Methods and results. In: Proceedings of the European Conference on Computer Vision (ECCV) Workshops (2024)

\bibitem{aim2024snr_dataset}
Nazarczuk, M., Tanay, T., Catley-Chandar, S., Shaw, R., Timofte, R., Pérez-Pellitero, E.: {AIM} 2024 sparse neural rendering challenge: Dataset and benchmark. In: Proceedings of the European Conference on Computer Vision (ECCV) Workshops (2024)

\bibitem{nuutinen2016cvd2014}
Nuutinen, M., Virtanen, T., Vaahteranoksa, M., Vuori, T., Oittinen, P., H{\"a}kkinen, J.: Cvd2014—a database for evaluating no-reference video quality assessment algorithms. IEEE Transactions on Image Processing  \textbf{25}(7),  3073--3086 (2016)

\bibitem{paudyal2014study}
Paudyal, P., Battisti, F., Carli, M.: A study on the effects of quality of service parameters on perceived video quality. In: 2014 5th European Workshop on Visual Information Processing (EUVIP). pp.~1--6. IEEE (2014)

\bibitem{haar}
Reisenhofer, R., Bosse, S., Kutyniok, G., Wiegand, T.: A haar wavelet-based perceptual similarity index for image quality assessment. Signal Processing Image Communication  \textbf{61},  33--43 (02 2018). \doi{10.1016/j.image.2017.11.001}

\bibitem{sinno2018large}
Sinno, Z., Bovik, A.C.: Large-scale study of perceptual video quality. IEEE Transactions on Image Processing  \textbf{28}(2),  612--627 (2018)

\bibitem{stroud2020d3d}
Stroud, J., Ross, D., Sun, C., Deng, J., Sukthankar, R.: D3d: Distilled 3d networks for video action recognition. In: Proceedings of the IEEE/CVF Winter Conference on Applications of Computer Vision. pp. 625--634 (2020)

\bibitem{wang2016mcl}
Wang, H., Gan, W., Hu, S., Lin, J.Y., Jin, L., Song, L., Wang, P., Katsavounidis, I., Aaron, A., Kuo, C.C.J.: Mcl-jcv: a jnd-based h. 264/avc video quality assessment dataset. In: 2016 IEEE International Conference on Image Processing (ICIP). pp. 1509--1513. IEEE (2016)

\bibitem{wang2017videoset}
Wang, H., Katsavounidis, I., Zhou, J., Park, J., Lei, S., Zhou, X., Pun, M.O., Jin, X., Wang, R., Wang, X., et~al.: Videoset: A large-scale compressed video quality dataset based on jnd measurement. Journal of Visual Communication and Image Representation  \textbf{46},  292--302 (2017)

\bibitem{clip}
Wang, J., Chan, K.C., Loy, C.C.: Exploring clip for assessing the look and feel of images. In: AAAI (2023)

\bibitem{wang2019youtube}
Wang, Y., Inguva, S., Adsumilli, B.: Youtube ugc dataset for video compression research. In: 2019 IEEE 21st International Workshop on Multimedia Signal Processing (MMSP). pp.~1--5. IEEE (2019)

\bibitem{wang2021rich}
Wang, Y., Ke, J., Talebi, H., Yim, J.G., Birkbeck, N., Adsumilli, B., Milanfar, P., Yang, F.: Rich features for perceptual quality assessment of ugc videos. In: Proceedings of the IEEE/CVF Conference on Computer Vision and Pattern Recognition. pp. 13435--13444 (2021)

\bibitem{msssim}
Wang, Z., Simoncelli, E.P., Bovik, A.C.: Multiscale structural similarity for image quality assessment. In: The Thrity-Seventh Asilomar Conference on Signals, Systems \& Computers, 2003. vol.~2, pp. 1398--1402. Ieee (2003)

\bibitem{wu2022fast}
Wu, H., Chen, C., Hou, J., Liao, L., Wang, A., Sun, W., Yan, Q., Lin, W.: Fast-vqa: Efficient end-to-end video quality assessment with fragment sampling. In: European conference on computer vision. pp. 538--554. Springer (2022)

\bibitem{wu2023qinstruct}
Wu, H., Zhang, Z., Zhang, E., Chen, C., Liao, L., Wang, A., Xu, K., Li, C., Hou, J., Zhai, G., Xue, G., Sun, W., Yan, Q., Lin, W.: Q-instruct: Improving low-level visual abilities for multi-modality foundation models (2023)

\bibitem{wu2023qalign}
Wu, H., Zhang, Z., Zhang, W., Chen, C., Li, C., Liao, L., Wang, A., Zhang, E., Sun, W., Yan, Q., Min, X., Zhai, G., Lin, W.: Q-align: Teaching lmms for visual scoring via discrete text-defined levels. arXiv preprint arXiv:2312.17090  (2023)

\bibitem{ying2021patch}
Ying, Z., Mandal, M., Ghadiyaram, D., Bovik, A.: Patch-vq:'patching up'the video quality problem. In: Proceedings of the IEEE/CVF Conference on Computer Vision and Pattern Recognition. pp. 14019--14029 (2021)

\bibitem{paq2piq}
Ying, Z., Niu, H., Gupta, P., Mahajan, D., Ghadiyaram, D., Bovik, A.: From patches to pictures (paq-2-piq): Mapping the perceptual space of picture quality. In: Proceedings of the IEEE/CVF Conference on Computer Vision and Pattern Recognition. pp. 3575--3585 (2020)

\bibitem{HVS_5M}
Zhang, A.X., Wang, Y.G., Tang, W., Li, L., Kwong, S.: A spatial–temporal video quality assessment method via comprehensive hvs simulation. IEEE Transactions on Cybernetics  \textbf{54}(8),  4749--4762 (2024). \doi{10.1109/TCYB.2023.3338615}

\bibitem{lpips}
Zhang, R., Isola, P., Efros, A.A., Shechtman, E., Wang, O.: The unreasonable effectiveness of deep features as a perceptual metric. In: 2018 IEEE/CVF Conference on Computer Vision and Pattern Recognition. pp. 586--595 (2018). \doi{10.1109/CVPR.2018.00068}

\bibitem{dbcnn}
Zhang, W., Ma, K., Yan, J., Deng, D., Wang, Z.: Blind image quality assessment using a deep bilinear convolutional neural network. IEEE Transactions on Circuits and Systems for Video Technology  \textbf{30}(1),  36--47 (2020)

\bibitem{zhang2023blind}
Zhang, W., Zhai, G., Wei, Y., Yang, X., Ma, K.: Blind image quality assessment via vision-language correspondence: A multitask learning perspective. In: Proceedings of the IEEE/CVF conference on computer vision and pattern recognition. pp. 14071--14081 (2023)

\bibitem{meta}
Zhu, H., Li, L., Wu, J., Dong, W., Shi, G.: {MetaIQA:} deep meta-learning for no-reference image quality assessment. In: Proceedings of the IEEE Conference on Computer Vision and Pattern Recognition (CVPR). pp. 14143--14152 (Jun 2020)

\end{thebibliography}
\end{document}